\documentclass[12pt]{elsarticle}
\usepackage{graphicx}
\usepackage{dcolumn}
\usepackage{bm}
\usepackage{mciteplus}
\usepackage{mathrsfs}
\usepackage{amsmath}

\newcommand{\dd}{{\! \! \rm d}}
\newcommand{\rr}{{\bf r}}

\newcommand{\RR}{{\bf R}}

\newcommand{\beq}{\begin{eqnarray}}
\newcommand{\eeq}{\end{eqnarray}}

\newcommand{\no}{\nonumber \\}
\newcommand{\ba}{\bar a}
\newcommand{\bl}{\bar \ell}

\journal{Physica A}

\begin{document}
\bibliographystyle{apsrevM}

\begin{frontmatter}

\title{Effective Hamiltonian for a liquid-gas interface fluctuating around a corrugated cylindrical substrate in the presence of van der Waals interactions}
\author{F. Dutka and M. Napi\'orkowski}
\address{Instytut Fizyki Teoretycznej, Uniwersytet Warszawski, \\ 00-681 Warszawa, Ho\.za 69, Poland}

\begin{abstract}
We investigate liquid layers adsorbed at spherical and corrugated cylindrical substrates. The effective Hamiltonians for the 
liquid-gas interfaces fluctuating in the presence of such curved substrates are derived via the mean-field density functional theory. 
Their structure is compared with the Helfrich Hamiltonian which is parametrized by the bending and 
Gaussian rigidity coefficients. For long-ranged interparticle interactions of van der Waals type these coefficients turn out to be non-universal functions of interfacial curvatures; their form varies from one interface to another. We discuss implications of the structure of these functions on the effective Hamiltonian. 
\end{abstract}


\end{frontmatter}

\section{Introduction}
The morphology of wetting layers adsorbed at planar substrates is nowadays rather well understood \cite{Rowlinson2,Dietrich1,Schick1,Henderson1}. 
However, adsorption at non-planar substrates is rather ubiquitous in nature and one is led to study the behavior of inhomogeneous fluids 
in the presence of curved substrates, for example the adsorption at large colloidal particles \cite{Holyst1,Stewart1,Upton1} 
and fibers \cite{Henderson2,Darbellay1,Gelfand1}, capillary condensation in porous media, liquid bridge formations between spheres \cite{Bauer2,Dobbs1,*Dobbs2} and cylinders \cite{Bauer1,Osborn1}. Though 60 years passed since the landmark work of 
Tolman \cite{Tolman1} on the surface tension of curved interfaces there are still many open problems related to adsorption at 
curved substrates. 

Two basic approaches to the problem of the surface tension coefficient dependence on interfacial curvature have been developed \cite{Blokhuis1}. The first 
is based on the analysis of the interface fluctuating around the planar configuration. 
The second approach focuses on the average shapes of the fluid interfaces curved around the non-planar substrates. In both cases the surface 
free-energy density is expanded in powers of curvature. The zeroth order term represents the surface tension coefficient of a planar interface, the 
coefficient in front of the linear term is related to the so-called Tolman length \cite{Tolman1} while the second order terms contain the bending and 
Gaussian rigidity coefficients. These two coefficients for membranes parametrize the phenomenological Helfrich Hamiltonian \cite{Helfrich1}. One of the still open problems in this field  is concerned with the status of this expansion and the existence of the bending and Gaussian rigidity coefficients. In particular 
one would like to know under what circumstances, if any,  one might expect the non-analytic 
dependence of the surface free-energy density on the interfacial curvature. 

In this paper we discuss these issues while investigating the fluctuating, cylindrically shaped liquid-gas interface. 
First (Section \ref{sect_sphere}) we follow the standard 
approach and calculate the sum of bending and Gaussian rigidity coefficients for the spherical geometry. Then we investigate the cylindrical, 
fluctuating interface (Section \ref{sect_cylinder}) and obtain information separately on the 
bending and on the Gaussian rigidity coefficients. We discuss the structure of the ensuing coefficients from the point of view of their dependence on the 
interparticle interaction and the substrate geometries (Section \ref{sect_rigidities}). In the last, Section \ref{summary}, we summarize our results.  

\section{The model}
In order to calculate the effective interface Hamiltonian for an interface separating a liquid-like layer adsorbed on the curved substrate from the gas phase we employ the density functional theory (DFT). The grand canonical
density functional $\Omega ([\rho(\rr)],T,\mu)$ depends parametrically on temperature $T$,  chemical potential $\mu$, interparticle potential $\tilde w(r)$ assumed to be spherically symmetric,  and on the external potential $V(\rr)$ representing the effect of the substrate. The interparticle potential $\tilde w(r)$ is split into the short-ranged repulsive part $w_{hs}(r)$ and the long-ranged attractive part $w(r)$ which is of the van der Waals type
\beq
	\tilde w(r) &=& w_{hs}(r)+w(r) \ .
\eeq  
The mean-field version of the density functional takes the form \cite{Evans2,Napiorkowski1,*Napiorkowski2,*Napiorkowski3}
\begin{align}
 \begin{split}
 \Omega ([\rho(\rr)],T,\mu) =&  \int \dd^3 r  f_{hs}(\rho(\rr)) + \frac{1}{2} \int \dd^3 r \!\! \int \dd^3 r' w(|\rr-\rr'|) \rho(\rr) \rho(\rr') \\
 	& + \int \dd^3 r \left( V_{ext}(\rr)-\mu \right)\rho(\rr) \, , \label{functional}
\end{split}
\end{align}
where the first term on the rhs represents the free energy of the fluid interacting via short-ranged repulsive potential $w_{hs}(r)$ evaluated in the local density approximation. In the following analysis the long-ranged attractive part of the potential $w(r)$ is  modeled by 
\beq 
 w(r) &=& - \frac{A}{(\kappa^2+r^2)^3} \  \label{potential}
\eeq
which decays $\sim r^{-6}$ at large distances (we neglect the retardation effect). The parameter $\kappa$ corresponds to the hard core radius of the fluid particles and the amplitude $A>0$ characterizes the strength of the attraction. The external potential $V_{ext}(\rr)$ acting on the fluid particle located at position $\rr$ comes from its interactions with all substrate particles. This interaction can be again split into the short- and long-ranged part.  The long-ranged part is again modeled by the potential $w_s(r)=-A_s/(\kappa_s^2+r^2)^3$, i.e. 
\beq \label{V_ext}
V_{ext}(\rr) &=& \int_{{\cal V}_s} \dd \rr' \, \rho_s \, w_s(|\rr-\rr'|) \ , 
\eeq 
where ${\cal V}_s$ denotes the region occupied by the homogeneous substrate with density $\rho_s$. The effect of the short-ranged repulsive part of the  substrate-fluid interaction is to prevent the fluid particles from penetrating the region ${\cal V}_s$; it is taken into account by the appropriate specification of the different regions integration present in the density functional, Eq.\,(\ref{functional}).

The thermodynamic state of the fluid is taken to be slightly off the bulk coexistence line in the regime of the stable bulk gas phase, and 
far away from the critical point. This implies that the bulk correlation length is comparable with the size of fluid particle  characterized by $\kappa$. In such circumstances the position of the liquid-gas interface is represented by function $z=f(\RR)$, $\RR=(x,y)$, and the nonuniform density profile $\rho(\rr)$ can be described by the so called sharp-kink approximation 
\beq 
 \rho_{shk}({\bf R},z) = \rho_{l}\,\Theta(f({\bf R})-z)\, + \, \rho_{g}\,\Theta(z-f({\bf R})) \ , \label{profile}
\eeq 
where $\Theta(z)$ is the Heaviside function while $\rho_{l}$ and $\rho_{g}$ denote the densities of the coexisting bulk liquid and gas phases, respectively. For a finite system the density functional $\Omega ([\rho(\rr)],T,\mu)$, Eq.\,(\ref{functional}), evaluated at 
$\rho(\rr) = \rho_{shk}({\bf R},z)$ can be represented as the sum of bulk, surface, line, etc. contributions  
 \cite{Napiorkowski1,*Napiorkowski2,*Napiorkowski3}. We discuss two shapes of substrates which induce the corresponding shapes 
 of interfaces: a spherical substrate of radius $R_s$, and an axially symmetric substrate represented by a corrugated cylinder of infinite length. In each case the gas phase is present away from the substrate while a liquid-like layer is adsorbed on the substrate. Our analysis is concentrated on surface and interfacial contributions to $\Omega ([\rho(\rr)],T,\mu)$. 


\section{Spherical interface \label{sect_sphere}}

The system under consideration in this section consists of the spherical substrate of radius $R_s$ on which a uniform liquid layer of constant thickness $\ell$ is adsorbed. The remaining part of the system is filled with the gas phase of volume 
$V_g$ while the volume of the liquid layer is denoted by $V_l$. The density functional, Eq.\,(\ref{functional}), evaluated for this geometry and for the spherically symmetric density profile in the sharp-kink approximation - after taking into account contributions due to the finite size of the system - equals (the upper index ${(s)}$ refers to the spherical case and the lower index $s$ refers to the surface properties)
\beq
 \Omega^{(s)} &=& (V_g+V_l) \, \omega_b(\rho_g) + \Omega^{(s)}_s \quad,
\eeq
where $\omega_b(\rho)$ is the grand canonical potential density of a fluid system with the uniform density $\rho$
\beq
\omega_b(\rho) = f_{hs}(\rho)+\frac{1}{2} \ \rho^2 \int \dd \rr \, w(|\rr|) - \mu \rho \quad.
\eeq 
The surface contribution $\Omega_s^{(s)}$ has the following form 
\begin{align}  \label{surface}
 \Omega_s^{(s)}= 4 \pi R_s^2 \omega_s^{(s)} = 4 \pi R_s^2 \left( \omega^{(s)}_{ex} + \sigma^{(s)}_{sl}(R_s) + 
    \frac{(R_s+\ell)^2}{R_s^2} \sigma^{(s)}_{lg}(R_s+\ell) + \omega^{(s)}_{int}(R_s,\ell) \right) .
\end{align}
The first term on the rhs in Eq.\,(\ref{surface}) is proportional to the volume of adsorbed liquid layer of thickness $\ell$
\beq
 \omega^{(s)}_{ex} &=& \frac{(R_s+\ell)^3-R_s^3}{3R_s^2} \, \omega_b(\rho_l) \quad.
\eeq
The next two terms $\sigma^{(s)}_{sl}(R_s)$ and $\sigma^{(s)}_{lg}(R_s+\ell)$ are the surface tension coefficients of the substrate-liquid and liquid-gas spherical interfaces with radii $R_s$ and $R_s+\ell$, respectively. The last term describes the interaction between the substrate-liquid and liquid-gas interfaces 
\beq
\label{int1}
\omega^{(s)}_{int}(a,\ell) &=& \Delta \rho \Big( \rho_l \,  \hat \omega^{(s)}_{int}(a,\ell,[w]) - \rho_s \,  
  \hat \omega^{(s)}_{int}(a,\ell,[w_s]) \Big) \quad,
\eeq
where $\Delta \rho = \rho_l-\rho_g$, and 
\beq
\hat \omega^{(s)}_{int}(a,\ell,[w])  &=& \frac{1}{4 \pi a^2} \int_{|\rr| \geq a+\ell} \dd^3 r   \int_{|\rr| \leq a} \dd^3 r' \, w(|\rr-\rr'|) \quad.
\eeq
Within the sharp-kink approximation employed in this paper each of the two contributions to the interaction between the substrate-liquid and liquid-gas interfaces, Eq.(\ref{int1}), factorizes into the term depending on the densities only, and the term determined exclusively by the interparticle interaction and geometry; the latter one is denoted with a hat. With its help one can express all the interface-interface interactions and surface tension coefficients in the problem at hand. For  example, the expressions for the surface tension coefficients take the following form
\beq
\sigma^{(s)}_{sl}(a) &=& -\frac{1}{2} \rho_l \Big( \rho_l \, \hat \omega^{(s)}_{int}(a,0,[w]) 
  - 2 \rho_s \, \hat \omega^{(s)}_{int}(a,0,[w_s]) \Big) \ ,
\eeq
\beq
 \sigma^{(s)}_{lg}(a) &=& -\frac{1}{2} (\Delta \rho)^2 \hat \omega^{(s)}_{int}(a,0,[w]) \ . \label{sigma_lg} 
\eeq
In the limit of large sphere radii one obtains the well known expression \cite{Napiorkowski1,*Napiorkowski2,*Napiorkowski3}
for the surface tension coefficient for the planar liquid-gas interface $\sigma_{lg}^{(p)}$
\beq
 \sigma^{(s)}_{lg}(a \to \infty) = \sigma_{lg}^{(p)}= -\frac{1}{2} (\Delta \rho)^2 \, \pi \int_0^\infty \dd  r \, r^3 w(r)
 \equiv -\frac{1}{2} (\Delta \rho)^2 \hat \sigma_{lg}^{(p)} \ .
\eeq

For the van der Waals interactions, Eqs\,(\ref{potential},\ref{V_ext}) one gets 
\cite{Stewart1,Bieker1}
\begin{align}
 \begin{split}
 \hat \omega^{(s)}_{int}(a,\ell,[w]) =& \ \hat \sigma^{(p)}_{lg} \frac{1}{6 \ba^2} \left\{4 \ba (\ba+\bl)+4 \ba^3 \arctan \frac{1}{2\ba+\bl} 
 	\right. \\
 &  -2 \bl(3\ba^2+3\ba \bl+\bl^2)\arctan \frac{2\ba}{1+\bl(2\ba+\bl)} \\
 & \left. +\ln(1+\bl^2)-\ln[1+(2\ba+\bl)^2]\right\} \ , 
\end{split}
\end{align}
where parameters $\ba = a/\kappa$, $\bl=\ell/\kappa$ are dimensionless, and 
\beq
	\hat \sigma^{(p)}_{lg} \,=\, - \frac{A \pi}{4 \kappa^2} \ < 0 \ .
\eeq

Similarly, the interaction between the spherical substrate-liquid and the spherical liquid-gas interfaces tends in the limit $1/\ba \to 0$, 
$\bar \ell$ fixed to the interaction between two planar interfaces 
\beq \label{omega_p}
	\hat \omega^{(p)}_{int}(\ell,[w]) &=&\hat \sigma^{(p)}_{lg} \left[ 1-\bl \arctan (1/\bl) \right] \quad.
\eeq
The approach of $\hat \omega^{(s)}_{int}(a,\ell,[w])$ to $\hat \omega^{(p)}_{int}(\ell,[w])$ is described by the following 
asymptotic expression
\beq
  \hat \omega^{(s)}_{int}(a,\ell,[w]) =  \left[1+ \frac{\ell}{a}+\ldots \right] \hat \omega^{(p)}_{int}(\ell,[w]) \quad.
\eeq

Using Eq.\,(\ref{sigma_lg}) for the van der Waals interactions one gets the following expression for the liquid-gas interfacial tension coefficient of spherical interface with radius $a$
\beq
 \hat \sigma^{(s)}_{lg}(a) &=& \hat \sigma^{(p)}_{lg} \frac{2}{3} \left[1+\ba \arctan(1/2 \ba)-\frac{\ln (1+4\ba^2)}{4\ba^2} \right] \ ,
\eeq
In the limit of small curvature $1/\ba$ the above expression takes the asymptotic form
\beq
  \hat \sigma^{(s)}_{lg}(a) = \hat \sigma^{(p)}_{lg} \left[1+\frac{1}{36}\left( -1-12 \ln 2 + 12 \ln \frac{1}{\ba} \right)\frac{1}{\ba^2}+\ldots \right] \quad.
  \label{sigma_s}
\eeq
The interfacial tension coefficient $\hat \sigma^{(s)}_{lg}(a)$ is a non-analytic function of dimensionless curvature $1/\ba$. It is 
smaller than in the planar case $\hat \sigma^{(p)}_{lg}$ and attains the planar limit from below. The first curvature dependent correction to the planar surface tension coefficient is proportional to $\ln(1/\ba) / \ba^2$. Qualitatively similar results have been obtained also for a different model of a long-ranged interparticle potential \cite{Stewart1,Bieker1}.

\section{Local Hamiltonian \label{sect_local}}
Fluctuating membranes can be analyzed with the help of a local Hamiltonian proposed by Helfrich \cite{Helfrich1}. It 
represents the energy of a membrane in terms of the mean $H=(1/R_1+1/R_2)/2$ and Gaussian $G=1/R_1\,R_2$ curvatures, 
where $R_1$ and $R_2$ denote the principal radii of local curvature. Hamiltonian of similar form has been also derived 
for fluctuating liquid-gas interfaces \cite{Napiorkowski1,*Napiorkowski2,*Napiorkowski3} 
\beq
 \mathscr{H}_{Helf} &=& \int \dd^2 s \left[\sigma^{(p)}_{lg} + k_H (H-C_0)^2+k_G\,G  \right] \quad, \label{Helfrich} 
\eeq
where $\ \dd^2 s$ denotes the surface element, $C_0$ - the spontaneous curvature, and $k_H$, $k_G$ are the coefficients 
of  bending and Gaussian rigidity, respectively. The parameter $C_0$ measures the profile assymetry \cite{Robledo1,Rochin2,Blokhuis3} and for the sharp-kink approximation used throughout this paper it is zero. The first non-vanishing corrections to the planar interface Hamiltonian are quadratic in inverse curvature radii. Accordingly, for the spherical liquid-gas interface with radius $a$ the integrand in Eq.\,(\ref{Helfrich})  can be written as 
\beq
\sigma_{lg}^{(s)}(a) &=& \sigma^{(p)}_{lg} + k^{(s)}_H \, H^2+k^{(s)}_G \, G \quad, 
\eeq
where $H^2 = G = 1/a^2$. 
Comparison of the above equation with Eq.(\ref{sigma_s}) shows that the sum of bending and Gaussian rigidity coefficients depends on the curvature radius
\beq
\label{sumkHkG}
k^{(s)}_H(a) +k^{(s)}_G(a) &=& \sigma^{(p)}_{lg} \kappa^2 \frac{1}{36}\left( -1-12 \ln 2 +12 \ln \frac{1}{\ba} \right) \ .
\eeq
The logarithmic dependence of the sum in Eq.(\ref{sumkHkG}) on the radius of curvature reflects the presence of 
van der Waals interparticle interactions in the system under consideration \cite{Stewart1,Bieker1}.


\section{Fluctuating cylindrical interface \label{sect_cylinder}} 
\begin{figure}[htb]
 \begin{center}
   \includegraphics{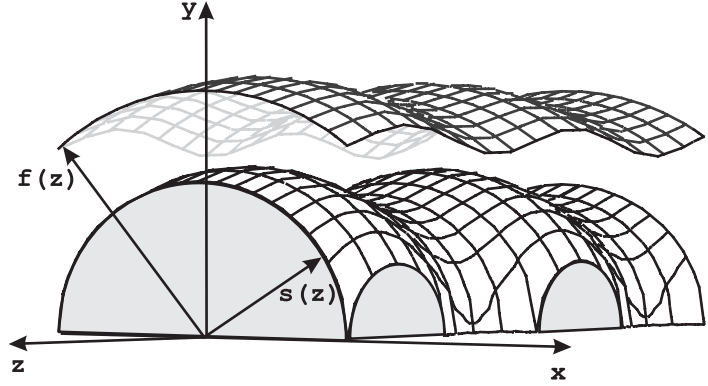}
   \caption{Section of the system with  axially symmetric substrate $r_\bot=s(z)$ and the liquid-gas interface $r_\bot=f(z)$.  \label{fig_cylinder}}
 \end{center}
\end{figure}

In this section we analyze the liquid-gas interface which fluctuates around its cylindrical configuration. This fluctuating interface separates the liquid layer adsorbed at the corrugated cylindrical substrate from the gas phase. For simplicity we assume 
that both the fluctuating liquid-gas interface and the corrugated substrate surface have axial symmetry and - in cylindrical 
coordinates $r_\bot, \phi, z$ - their positions are given by $r_\bot=f(z)$, and $r_\bot=s(z)$, respectively;  
see Fig.\,\ref{fig_cylinder}. We aim at deriving the effective Hamiltonian $\mathscr{H}_{eff}^{(c)}[f]$ for this fluctuating liquid-gas
interface. After evaluating the density functional $\Omega ([\rho(\rr)],T,\mu)$  for the sharp-kink density profile 
one extracts the surface contribution \linebreak $\Omega_s^{(c)}([s],[f],T,\mu)$ to $\Omega ([\rho(\rr)],T,\mu)$. It is a functional of both the 
substrate surface shape and the shape of the fluctuating interface (the dependence on thermodynamic parameters is not displayed) 
\beq
  \Omega_s^{(c)}([s],[f]) &=& \Omega_{sl}^{(c)}([s])+\Omega_{lg}^{(c)}([f])+\Omega^{(c)}_{int}([s],[f]) \quad.
\eeq
The first two terms correspond to the free-energy functionals of the corrugated solid-liquid and the liquid-gas interfaces, 
respectively. The last term denotes the interaction between these two interfaces  
\beq
 \Omega^{(c)}_{int}([s],[f]) &=& \Delta \rho \left(\rho_l \hat \Omega^{(c)}_{int}([s],[f],[w]) - \rho_s \hat \Omega^{(c)}_{int}([s],[f],[w_s])  \right) \ , \label{int_c}
\eeq
where
\begin{align}
\hat \Omega^{(c)}_{int}([s],[f],[w]) = \int \dd z \int \dd z' \int_{|\rr_\bot| \leq s(z)} \dd^2 r_\bot \int_{|\rr'_\bot| \geq f(z') }\dd^2 r'_\bot\, w (|\rr - \rr'|) \ .
\end{align}
The two-dimensional vectors $\rr_\bot$ and $\rr'_\bot$ are perpendicular to the $z$ axis, \mbox{$\rr = (\rr_\bot,z)$}. The range of integration over 
$z$ and $z'$ is determined by the size of the system along the symmetry axis, say $[-L,L]$. However, we are not interested in the finite size effects due to this cut-off along the $z$ axis and extend the integration limits $L \rightarrow \infty$ keeping in mind, that only the quantity $\hat \Omega^{(c)}_{int}([s],[f])  / L$ is well defined in this limit. 

Assuming small corrugation $|s'(z)| \ll 1$ one can perform the gradient expansion of $\hat \Omega^{(c)}_{int}([s],[f],[w]) $ up to bilinear terms (Appendix \ref{app}) and obtain the local form of the interaction energy
\begin{align}
 \begin{split}
  \hat \Omega^{(c)}_{int}([s],[f],[w]) = 2\pi \int \dd z \, s(z)  \Big\{  & \hat \omega_{int}^{(c)}\Big( s(z),f(z) \Big) \\
   	 	&  + \frac{s'(z) f'(z)}{2} \ \hat\lambda^{(c)}\Big(s(z),f(z)\Big) \Big\} \quad. 
\end{split}
\end{align}
The first term on the rhs describes the interaction between two coaxial, undulated cylindrical interfaces in the Derjaguin approximation \cite{Derjaguin1,*Goetzelmann1,*Troendle1}. The function $\hat \omega_{int}^{(c)}(a,b)$ itself describes the interaction of cylindrical interfaces of constant radii $a$ and $b$ ($a \leq b$), i.e. without any undulation. The second term is the correction due to undulation; it vanishes for ideally cylindrical shape $s(z)=const.$  This term turns out to be crucial in determining the coefficient of surface tension. 

\begin{figure}[htb]
 \begin{center}
   \includegraphics{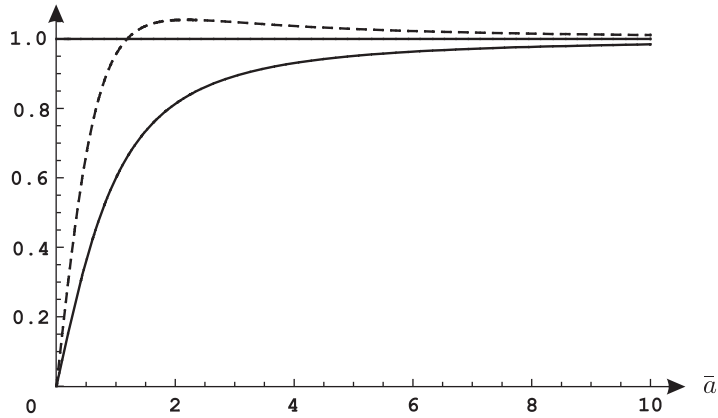} 
   \caption{The radius of curvature dependence of the functions $\hat \sigma^{(c)}_{lg}(a)$ (solid line) and $\hat \sigma^{(x)}_{lg}(a)$ (dotted line) in the $\hat \sigma^{(p)}_{lg}$ units. \label{fig_sigmax}}
 \end{center}
\end{figure}

The surface free energy of corrugated, axially symmetric substrate-liquid interface with $z$-dependent radius $\rr_\bot=s(z)$ is equal
\beq
 \Omega^{(c)}_{sl}([s]) &=& -\frac{1}{2} \rho_l \left( \rho_l \, \hat \Omega^{(c)}_{int}([s],[s],[w])-2 \rho_s \, \hat \Omega^{(c)}_{int}([s],[s],[w_s]) \right) \ ,
\eeq
while the corresponding expression for the surface free energy of liquid-gas interface of radius $r_\bot=f(z)$ is equal
\beq
 \Omega^{(c)}_{lg}([f]) &=& -\frac{1}{2} (\Delta \rho)^2 \hat \Omega^{(c)}_{int}([f],[f],[w]) \ .
\eeq
For small undulations $|f'(z)| \ll 1$ one can rewrite $\hat \Omega^{(c)}_{int}([f],[f],[w])$ in the local form 
\beq
 \hat \Omega^{(c), loc}_{lg}([f]) &=&  2\pi \int \dd z \, f(z) \left\{ \hat \sigma_{lg}^{(c)}(f(z))+ \frac{f'(z)^2}{2} \ \hat \sigma_{lg}^{(x)}(f(z)) \right\} \ , \label{local_form}
 \label{Omega_loc}
\eeq
where
\begin{align}
\begin{split}
\label{sigmas1}
 \hat \sigma_{lg}^{(c)}(a) &= \hat \omega_{int}^{(c)}(a,a) = \hat \sigma^{(p)}_{lg} \, \frac{1}{2 \ba} \, \Big( E(-4\ba^2)-K(-4 \ba^2)\Big) \\
 \hat \sigma_{lg}^{(x)}(a) &= \hat\lambda^{(c)}(a,a) = \hat \sigma^{(p)}_{lg} \, 2 \ba \,  \frac{ E (-4 \ba^2)}{1+4 \ba^2} \ .
\end{split}
\end{align}
The symbols $K(x)$ and $E(x)$ in Eq.(\ref{sigmas1}) denote the complete elliptic integrals of first and second kind \cite{Abramowitz1}, respectively. The expression $\hat \sigma_{lg}^{(c)}(a)$ represents the surface free energy of cylindrical surface of radius $a$ and is often shortly denoted as the $a$-dependent coefficient of surface tension. It is an increasing function of radius $a$, see Fig. \ref{fig_sigmax}. On the other hand, the coefficient in front of $\frac{1}{2}f'(z)^2$ in Eq.(\ref{Omega_loc}) is different from $\hat \sigma_{lg}^{(c)}$ and is denoted as 
$\hat \sigma_{lg}^{(x)}(a)$; it is not a monotonous function of $a$, see Fig. \ref{fig_sigmax}. 
It has a single maximum and for $a \to \infty$ it tends - similarly to $\hat \sigma^{(c)}_{lg}(a)$ - to the coefficient of the surface tension of planar interface 

\begin{align}
\begin{split}  \label{sigmas}
\hat \sigma_{lg}^{(c)}(a) &=\hat \sigma_{lg}^{(p)}  \left[ 1+ \frac{1}{16} \left(1 - 6 \ln 2 + 2 \ln \frac{1}{\bar a}\right) \frac{1}{\bar a^2}+\ldots \right] \\
\hat \sigma_{lg}^{(x)}(a) &= \hat \sigma_{lg}^{(p)} 
	\left[ 1+ \frac{1}{16} \left( -3 + 6 \ln 2 -2 \ln \frac{1}{\bar a} \right) \frac{1}{\bar a^2}+\ldots \right] \quad .
\end{split}
\end{align}

We recall that the local capillary-wave Hamiltonian for an interface fluctuating around a planar configuration and whose position is described by  $f=f(x,y)$ is - in the limit of small interfacial undulations - given by \cite{Napiorkowski1,*Napiorkowski2,*Napiorkowski3,Diehl1}
\beq
\mathscr{H}^{(p)}_{cw}[f] &=&  \int \dd x \! \! \int \dd y \, \sigma^{(p)}_{lg} \left(1+\frac{1}{2}(\nabla f)^2\right) 
	=  \int \dd^2 s \, \sigma^{(p)}_{lg} \ , \label{H_cw}
\eeq
where $\ \dd^2 s$ denotes the area element of the interface. One might expect that the local form of the Hamiltonian for an interface fluctuating around a cylindrical configuration will display the structure similar to the one in Eq.\,(\ref{H_cw}), i.e. the coefficients in front of $1$ and $\frac{1}{2}(\nabla f)^2$ in Eq.\,(\ref{Omega_loc}) will be the same which would amount to replacing $\sigma^{(p)}_{lg} \to \sigma^{(c)}_{lg}(f)$ in Eq.\,(\ref{H_cw}). However, this is not the case because $\hat \sigma^{(x)}_{lg}(f) \neq \hat \sigma^{(c)}_{lg}(f)$, and the Eq. (\ref{Omega_loc}) cannot be rewritten in a form similar to Eq.\,(\ref{H_cw}). 

To discuss this issue from a somewhat different perspective we return to the Helfrich Hamiltonian (Eq.\,(\ref{Helfrich})). For an axially symmetric interface $r_\bot = f(z)$, the mean and Gaussian curvatures are given by
\begin{align}
 \begin{split}
 H &= \frac{1}{2} \left(\frac{d}{dz} \frac{f'}{\sqrt{1+f'^2}}-\frac{1}{f \sqrt{1+f'^2}} \right) \\
 G &= - \frac{f''}{f (1+f'^2)^2} \ , 
\end{split}
\end{align}
and the Helfrich Hamiltonian expanded up to terms quadratic in $f'(z)$ - after some algebra, integration by parts - is equal to 
\begin{align}
 \begin{split}
 \mathscr{H}_{Helf}^{(c)}[f] = 2 \pi \int \dd z \ f \left[\sigma^{(p)}_{lg}+\frac{k^{(c)}_H}{4 f^2}
 	+\left(\sigma^{(p)}_{lg}-\frac{k^{(c)}_H}{4 f^2} \right)\frac{f'^2}{2} \right.
 	  \\ \left.  +\frac{f'}{f} \frac{d}{dz} \left(\frac{k^{(c)}_H}{2}+k^{(c)}_G \right) \right] \quad, \label{Helf_f}
 \end{split}
\end{align}
where the last term indicates that we allow the rigidity coefficients $k^{(c)}_H, k^{(c)}_G$ to be considered as functions of the 
local interface position $f(z)$. 
Comparing the above expression with the local form of the surface free energy obtained within the density functional approach, Eq.\,(\ref{Omega_loc}), one  gets the following expressions for the bending and Gaussian rigidity coefficients
\begin{align}
 \begin{split}
 k^{(c)}_H(a) &= \sigma^{(p)}_{lg} \kappa^2 \frac{1}{4} \left(1 - 6 \ln 2 + 2 \ln \frac{1}{\ba} \right) \\
 k^{(c)}_G(a) &= -\sigma^{(p)}_{lg} \kappa^2 \frac{3}{16} \ln \frac{1}{\ba} + k_{0} \ . 
\end{split}
\end{align}
Note that the $\mathscr{H}_{Helf}^{(c)}[f]$ dependence on the Gaussian rigidity coefficient, Eq.\,(\ref{Helf_f}), is only via its derivative and thus the function $k^{(c)}_G$ is known only up to an arbitrary integration constant $k_{0}$. The value of this constant is irrelevant because the integral of the Gaussian curvature over entire interfacial area is zero on the basis of the Gauss--Bonnet theorem \cite{Kreyszig1}. The logarithmic terms in rigidity coefficients arise from the corresponding logarithmic terms in surface tension coefficients in Eq.\,(\ref{sigmas}).


\section{Bending and Gaussian rigidity coefficients \label{sect_rigidities}}
For short-ranged forces the mean and Gaussian rigidity coefficients which appear in the Helfrich Hamiltonian (Eq.\,(\ref{Helfrich})) are constant. They do 
not depend on the actual interfacial geometry, and are usually derived either by considering the thermal fluctuations around the planar configuration, or by investigating the mean shapes of interfaces curved around cylindrical and spherical substrates. In the first approach based on the density functional theory supplemented by the sharp-kink approximation these coefficients are obtained as the fifth moments of the attractive part of the interparticle interaction \cite{Napiorkowski1,*Napiorkowski2,*Napiorkowski3} 
\beq \label{rigidities_planar}
k_H^{(p)} &=&  \frac{ \pi}{16} (\Delta \rho)^2 \int_0^\infty \dd r \, r^5 \, w(r) \no
k_G^{(p)} &=& -\frac{1}{3} k_H^{(p)} \ , 
\eeq 
while the surface tension coefficient is represented by the third moment of this interaction
\beq
 \sigma_{lg}^{(p)} &=& - \frac{\pi}{2} (\Delta \rho)^2 \int_0^\infty \dd r \, r^3 \, w(r) \ . \label{sigma_planar} 
\eeq 
The second approach consists of two steps. First, from analyzing the cylindrical shape of the liquid-gas interface (with zero Gaussian curvature) one derives the expression for the bending rigidity coefficient. Second, from considering the spherical interface  (Section\,\ref{sect_sphere}) one gets the expression for the sum of bending and Gaussian rigidity coefficients. Assuming the geometry independence of these coefficients, one can obtain also the expression for the Gaussian rigidity coefficient \cite{Giessen2,Segovia1,Dean1,Blokhuis2}. 

The bending and Gaussian rigidity coefficients derived in this way are equal to those derived in the former approach, provided the fifth moment of attractive part of the potential exists. On the other hand in the case of van der Waals interactions the rigidity coefficients for cylindrical and spherical geometries turn into functions of the mean curvature of the interface \mbox{\cite{Stewart1,Bieker1,Segovia1,Dean1}}. 
This fact has been already noted in the literature. However, the bending and Gaussian rigidity coefficient functions were derived within the  second approach \cite{Segovia1,Dean1} by assuming that they depend on the interfacial curvature in a universal way, i.e. in the same way for the cylindrical and spherical geometry.  

By investigating the liquid-gas interface fluctuating around the corrugated cylindrical substrate we derived expressions for both the bending  and Gaussian rigidity coefficients.  The sum of these quantities differs from those obtained by considering the spherical interface. It means that for van der Waals forces the bending and Gaussian rigidity coefficients are geometry dependent. We stress that these conclusions apply  only for interparticle potentials for which the cylindrical Gaussian rigidity coefficient depends on the curvature of the interface, e.g. for potential in Eq.\,(\ref{potential}). Otherwise the Gaussian rigidity coefficient cannot be derived using this method because of the Gauss--Bonnet theorem. 

The above non-universal behavior of bending and Gaussian rigidity coefficients is restricted to the type of interparticle potential considered in this paper. Analogous considerations on the level of the liquid-gas surface tension coefficient for interparticle potentials with  $r^{-4}$ decay at large distances, e.g. $w(r)=-A/(\kappa^2+r^2)^2$, lead to the conclusion that already the surface tension coefficient in the planar geometry (Eq.\,(\ref{sigma_planar})) does not exist. The surface tension coefficients in cylindrical ($\sigma_{lg}^{(c)}(R_c)$) and spherical ($\sigma_{lg}^{(s)}(R_s)$) geometries can be expressed in terms of elliptic integrals and tend to infinity in the limit $R_{c,s} \to \infty$. The dominant terms take the following form 
\begin{align}
\begin{split}
\sigma_{lg}^{(c)}(R_c) - \left[- \frac{A \pi}{2}(\Delta \rho)^2 \left( 1-3\log2 + \log \frac{\kappa}{R_c} \right) \right]
	& \mathop {\longrightarrow}\limits_{R_c \to \infty}  0 \ , \\
\sigma_{lg}^{(s)}(R_s) - \left[ - \frac{A \pi}{2}(\Delta \rho)^2 \left( -\log2 + \log \frac{\kappa}{R_s} \right) \right]
	&  \mathop {\longrightarrow}\limits_{R_s \to \infty} 0 \ .
\end{split}
\end{align}
Thus also in this case one cannot define the curvature dependent surface tension coefficient which is universal in this sense that it does not distinguish from which geometry the curvature comes from.


\section{Summary \label{summary}}
The effective Hamiltonian for the liquid-gas interface fluctuating around the corrugated cylindrical substrate has been derived 
within the mean-field version of density functional theory for fluid whose particles interact via van der Waals potential. This Hamiltonian is non-local but in the limit of small interfacial undulations and small cylindrical substrate corrugation it takes 
local form.  However, the structure of the effective Hamiltonian, Eq.\,(\ref{local_form}), in its local form is qualitatively different from the one describing the fluctuating interface around the planar configuration Eq.\,(\ref{H_cw}); it cannot be presented as the integral of the local curvature dependent surface tension coefficient over the interfacial area. 

The local effective Hamiltonian for the fluctuating cylindrical interface can be rewritten in the form proposed by Helfrich only when the rigidity coefficients multiplying the mean curvature squared and the Gaussian curvature are replaced by functions of the local curvature. The structure of these functions turns out to be non-universal, i.e., their form depends on whether one considers spherical or cylindrical interface.

\pagebreak
\appendix
\section{The gradient expansion of the interaction energy between two corrugated cylindrical interfaces \label{app} }
The interaction energy of two axially symmetric interfaces $r_\bot=s(z)$ and $r_\bot=f(z)$, $f(z) > s(z)$  is equal 
\begin{align}
 \begin{split}
\hat \Omega^{(c)}_{int}([s],[f],[w]) =& \int \dd z \int \dd z' \int_{|\rr_\bot| \leq s(z)} \dd^2 r_\bot 
 	\int_{|\rr'_\bot| \geq f(z') }\dd^2 r'_\bot  w (|\rr - \rr'|) \\
 	=& \int \dd z \int \dd z' \left[ J^{(c)}(s(z),\infty,z-z')-J^{(c)}(s(z),f(z'),z-z') \right] \ ,
 \end{split}
\end{align}
where 
\beq
  J^{(c)}(a,b,z') &=& \int_{0 \leq |r_\bot| \leq a} \dd^2 r_\bot \int_{0 \leq |r'_\bot| \leq b}\dd^2 r'_\bot 
 		w \left(\sqrt{(\rr_\bot - \rr'_\bot)^2+z'^2}\right) 
\eeq
denotes the interaction between two coaxial discs with radii $a$ and $b$ separated by distance $z'$. Assuming small corrugations, 
i.e., $|f'(z)|, |s'(z)| \ll 1$ one obtains
\begin{align}
 \begin{split}
  \hat \Omega^{(c)}_{int}[s,f] =&  \int \dd z \int \dd z' \left[ J(s(z),\infty,z-z')-J(s(z'),f(z),z-z') \right]  \\
 	 =&  \int \dd z \int \dd z' \left[ J(s(z),\infty,z')-J(s(z'+z),f(z),z') \right]  \\
   =&  \int \dd z \int \dd z' \left[ J(s(z),\infty,z')-J(s(z)+ s'(z)z'+s''(z)\frac{z'^2}{2},f(z),z') \right] \\
   =&  \int \dd z \int \dd z' \left[J(s(z),\infty,z')-J(s(z),f(z),z')\right]  \\
   &	\ - \int \dd z \int \dd z' \frac{s'(z)^2}{2} z'^2  \left[  
   	\frac{\partial^2 J(s(z),f(z),z')}{\partial s(z)} - \frac{1}{s'(z)} \frac{d}{dz} 
   		 \frac{\partial J(s(z),f(z),z')}{\partial s(z)} \right]  \\
   	=&  2\pi \int \dd z  s(z) \hat \omega_{int}^{(c)}(s(z),f(z)) + \int \dd z \int \dd z' \frac{s'(z) f'(z)}{2} z'^2 
   		\frac{\partial^2 J(s(z),f(z),z')}{\partial s(z) \partial (f(z))} \\
   	=& 2\pi \int \dd z \, s(z) \left\{ \hat \omega_{int}^{(c)}\Big( s(z),f(z) \Big) 
   	 	+ \frac{s'(z) f'(z)}{2} \ \hat\lambda^{(c)}\Big(s(z),f(z)\Big) \right\}  \ . 
 \end{split}
\end{align}
In the last line, the first term in the bracket $\hat \omega_{int}^{(c)}(a,b)$ denotes the interaction between two cylindrical interfaces of radii $a$ and $b$, respectively
\cite{Bieker1}
\begin{align}
 \hat \omega_{int}^{(c)}(a,b) = \left(- \frac{A \pi}{4 \kappa^2} \right) 
 \int \dd z' \frac{\kappa}{2a(z'^2+1)^2}\left(\frac{m^2+n(z'^2+1)}{\sqrt{(z'^2+1)^2+2n(z'^2+1)+m^2}} -m\right) \label{omega_c}
\end{align}
where $m={\bar b}^2-\ba^2$, and $n={\bar b}^2+\ba^2$ ($\bar b = b/\kappa$, $\ba = a/\kappa$). The second term $\hat\lambda^{(c)}(a,b)$ describes the coupling between the corrugations
\beq
\hat \lambda^{(c)}(a,b) &=&  \left(- \frac{A \pi}{4 \kappa^2}\right) 
 	 \frac{ 2 \bar b E\left(- \frac{4\bar b \bar a}{(\bar b-\bar a)^2+1}\right)}{((\bar b+\bar a)^2+1) \sqrt{(\bar b-\bar a)^2+1} }  \label{lambda_c} \ .
\eeq
In the limit $\ba,\,\bar b \gg 1$, $\bl = \bar b-\ba$ fixed, both quantities tend to its planar values
\begin{align}
 \hat \omega_{int}^{(c)}(a,b) =& \  \hat \omega_{int}^{(p)}(\ell) \left(1+\frac{\ell}{2a}+\ldots\right)  \\
 \hat \lambda^{(c)}(a,b) =& \ \hat \lambda^{(p)}(\ell) \left(1+\frac{\ell}{2a}+\ldots\right) \ ,
\end{align}
where $\hat \omega_{int}^{(p)}(\ell)$ is given by Eq.\,(\ref{omega_p}), and
\begin{align}
 \hat \lambda^{(p)}(\ell) = \left(- \frac{A \pi}{4 \kappa^2}\right) \frac{1}{1+\ell^2} \ .
\end{align}
On the other hand, for $a$ fixed, and $\bl \gg 1$ one has
\beq
   \hat \lambda^{(c)}(a,b) = \hat \lambda^{(p)}(\ell) \left[1- 2 \ba \frac{1}{\bl} +\left(\frac{21}{4}\ba^2-\frac{1}{2} \right)\frac{1}{\bl^2} + \ldots \right]  \, 
\eeq
and from numerical computation one obtains 
\beq
 \hat \omega_{int}^{(c)}(a,b) = \hat \omega_{int}^{(p)}(\ell) \left[\frac{3\pi}{2}\frac{a}{\ell}-\frac{9\pi}{2}\left(\frac{a}{\ell}\right)^2 +\ldots\right]  \ .
\eeq

\pagebreak
\bibliography{bibliografia}

\ifx\mcitethebibliography\mciteundefinedmacro
\PackageError{apsrevM.bst}{mciteplus.sty has not been loaded}
{This bibstyle requires the use of the mciteplus package.}\fi
\begin{mcitethebibliography}{36}
\expandafter\ifx\csname natexlab\endcsname\relax\def\natexlab#1{#1}\fi
\expandafter\ifx\csname bibnamefont\endcsname\relax
  \def\bibnamefont#1{#1}\fi
\expandafter\ifx\csname bibfnamefont\endcsname\relax
  \def\bibfnamefont#1{#1}\fi
\expandafter\ifx\csname citenamefont\endcsname\relax
  \def\citenamefont#1{#1}\fi
\expandafter\ifx\csname url\endcsname\relax
  \def\url#1{\texttt{#1}}\fi
\expandafter\ifx\csname urlprefix\endcsname\relax\def\urlprefix{URL }\fi
\providecommand{\bibinfo}[2]{#2}
\providecommand{\eprint}[2][]{\url{#2}}

\bibitem[{\citenamefont{Rowlinson and Widom}(1982)}]{Rowlinson2}
\bibinfo{author}{\bibfnamefont{J.~S.} \bibnamefont{Rowlinson}}
  \bibnamefont{and} \bibinfo{author}{\bibfnamefont{B.}~\bibnamefont{Widom}},
  \emph{\bibinfo{title}{Molecular Theory of Capillarity}}
  (\bibinfo{publisher}{Oxford University}, \bibinfo{address}{London},
  \bibinfo{year}{1982})\relax
\mciteBstWouldAddEndPuncttrue
\mciteSetBstMidEndSepPunct{\mcitedefaultmidpunct}
{\mcitedefaultendpunct}{\mcitedefaultseppunct}\relax
\EndOfBibitem
\bibitem[{\citenamefont{Dietrich}(1988)}]{Dietrich1}
\bibinfo{author}{\bibfnamefont{S.}~\bibnamefont{Dietrich}}, in
  \emph{\bibinfo{booktitle}{Phase transitions and critical phenomena}}, edited
  by \bibinfo{editor}{\bibfnamefont{C.}~\bibnamefont{Domb}} \bibnamefont{and}
  \bibinfo{editor}{\bibfnamefont{J.~L.} \bibnamefont{Lebowitz}}
  (\bibinfo{publisher}{Academic Press}, \bibinfo{address}{London},
  \bibinfo{year}{1988}), vol.~\bibinfo{volume}{12}, p.~\bibinfo{pages}{1}\relax
\mciteBstWouldAddEndPuncttrue
\mciteSetBstMidEndSepPunct{\mcitedefaultmidpunct}
{\mcitedefaultendpunct}{\mcitedefaultseppunct}\relax
\EndOfBibitem
\bibitem[{\citenamefont{Schick}(1989)}]{Schick1}
\bibinfo{author}{\bibfnamefont{M.}~\bibnamefont{Schick}}, in
  \emph{\bibinfo{booktitle}{Proceedings of the Les Houches Summer School,
  Session XLVIII}}, edited by
  \bibinfo{editor}{\bibfnamefont{J.}~\bibnamefont{Charvolin}},
  \bibinfo{editor}{\bibfnamefont{J.~F.} \bibnamefont{Joanny}},
  \bibnamefont{and}
  \bibinfo{editor}{\bibfnamefont{J.}~\bibnamefont{Zinn-Justin}}
  (\bibinfo{publisher}{Elsevier}, \bibinfo{address}{Amsterdam},
  \bibinfo{year}{1989}), p. \bibinfo{pages}{415}\relax
\mciteBstWouldAddEndPuncttrue
\mciteSetBstMidEndSepPunct{\mcitedefaultmidpunct}
{\mcitedefaultendpunct}{\mcitedefaultseppunct}\relax
\EndOfBibitem
\bibitem[{\citenamefont{Henderson}(1986)}]{Henderson1}
\bibinfo{author}{\bibfnamefont{J.~R.} \bibnamefont{Henderson}}, in
  \emph{\bibinfo{booktitle}{Fluid Interfacial Phenomena}}, edited by
  \bibinfo{editor}{\bibfnamefont{C.~A.} \bibnamefont{Croxton}}
  (\bibinfo{publisher}{John Wiley \& Sons Ltd.}, \bibinfo{address}{London},
  \bibinfo{year}{1986}), p. \bibinfo{pages}{555}\relax
\mciteBstWouldAddEndPuncttrue
\mciteSetBstMidEndSepPunct{\mcitedefaultmidpunct}
{\mcitedefaultendpunct}{\mcitedefaultseppunct}\relax
\EndOfBibitem
\bibitem[{\citenamefont{Holyst and Poniewierski}(1987)}]{Holyst1}
\bibinfo{author}{\bibfnamefont{R.}~\bibnamefont{Holyst}} \bibnamefont{and}
  \bibinfo{author}{\bibfnamefont{A.}~\bibnamefont{Poniewierski}},
  \bibinfo{journal}{Phys. Rev. B} \textbf{\bibinfo{volume}{36}},
  \bibinfo{pages}{5628} (\bibinfo{year}{1987})\relax
\mciteBstWouldAddEndPuncttrue
\mciteSetBstMidEndSepPunct{\mcitedefaultmidpunct}
{\mcitedefaultendpunct}{\mcitedefaultseppunct}\relax
\EndOfBibitem
\bibitem[{\citenamefont{Stewart and Evans}(2005)}]{Stewart1}
\bibinfo{author}{\bibfnamefont{M.~C.} \bibnamefont{Stewart}} \bibnamefont{and}
  \bibinfo{author}{\bibfnamefont{R.}~\bibnamefont{Evans}},
  \bibinfo{journal}{Phys. Rev. E} \textbf{\bibinfo{volume}{71}},
  \bibinfo{pages}{011602} (\bibinfo{year}{2005})\relax
\mciteBstWouldAddEndPuncttrue
\mciteSetBstMidEndSepPunct{\mcitedefaultmidpunct}
{\mcitedefaultendpunct}{\mcitedefaultseppunct}\relax
\EndOfBibitem
\bibitem[{\citenamefont{Upton et~al.}(1989)\citenamefont{Upton, Indekeu, and
  Yeomans}}]{Upton1}
\bibinfo{author}{\bibfnamefont{P.~J.} \bibnamefont{Upton}},
  \bibinfo{author}{\bibfnamefont{J.~O.} \bibnamefont{Indekeu}},
  \bibnamefont{and} \bibinfo{author}{\bibfnamefont{J.~M.}
  \bibnamefont{Yeomans}}, \bibinfo{journal}{Phys. Rev. B}
  \textbf{\bibinfo{volume}{40}}, \bibinfo{pages}{666}
  (\bibinfo{year}{1989})\relax
\mciteBstWouldAddEndPuncttrue
\mciteSetBstMidEndSepPunct{\mcitedefaultmidpunct}
{\mcitedefaultendpunct}{\mcitedefaultseppunct}\relax
\EndOfBibitem
\bibitem[{\citenamefont{Henderson and Rowlinson}(1984)}]{Henderson2}
\bibinfo{author}{\bibfnamefont{J.~R.} \bibnamefont{Henderson}}
  \bibnamefont{and} \bibinfo{author}{\bibfnamefont{J.~S.}
  \bibnamefont{Rowlinson}}, \bibinfo{journal}{J. Phys. Chem.}
  \textbf{\bibinfo{volume}{88}}, \bibinfo{pages}{6484}
  (\bibinfo{year}{1984})\relax
\mciteBstWouldAddEndPuncttrue
\mciteSetBstMidEndSepPunct{\mcitedefaultmidpunct}
{\mcitedefaultendpunct}{\mcitedefaultseppunct}\relax
\EndOfBibitem
\bibitem[{\citenamefont{Darbellay and Yeomans}(1990)}]{Darbellay1}
\bibinfo{author}{\bibfnamefont{G.~A.} \bibnamefont{Darbellay}}
  \bibnamefont{and} \bibinfo{author}{\bibfnamefont{J.~M.}
  \bibnamefont{Yeomans}}, \bibinfo{journal}{J. Phys. A.: Math. Gen.}
  \textbf{\bibinfo{volume}{23}}, \bibinfo{pages}{5655}
  (\bibinfo{year}{1990})\relax
\mciteBstWouldAddEndPuncttrue
\mciteSetBstMidEndSepPunct{\mcitedefaultmidpunct}
{\mcitedefaultendpunct}{\mcitedefaultseppunct}\relax
\EndOfBibitem
\bibitem[{\citenamefont{Gelfand and Lipowsky}(1987)}]{Gelfand1}
\bibinfo{author}{\bibfnamefont{M.~P.} \bibnamefont{Gelfand}} \bibnamefont{and}
  \bibinfo{author}{\bibfnamefont{R.}~\bibnamefont{Lipowsky}},
  \bibinfo{journal}{Phys. Rev. B} \textbf{\bibinfo{volume}{36}},
  \bibinfo{pages}{8725} (\bibinfo{year}{1987})\relax
\mciteBstWouldAddEndPuncttrue
\mciteSetBstMidEndSepPunct{\mcitedefaultmidpunct}
{\mcitedefaultendpunct}{\mcitedefaultseppunct}\relax
\EndOfBibitem
\bibitem[{\citenamefont{Bauer et~al.}(2000)\citenamefont{Bauer, Bieker, and
  Dietrich}}]{Bauer2}
\bibinfo{author}{\bibfnamefont{C.}~\bibnamefont{Bauer}},
  \bibinfo{author}{\bibfnamefont{T.}~\bibnamefont{Bieker}}, \bibnamefont{and}
  \bibinfo{author}{\bibfnamefont{S.}~\bibnamefont{Dietrich}},
  \bibinfo{journal}{Phys. Rev. E} \textbf{\bibinfo{volume}{62}},
  \bibinfo{pages}{5324} (\bibinfo{year}{2000})\relax
\mciteBstWouldAddEndPuncttrue
\mciteSetBstMidEndSepPunct{\mcitedefaultmidpunct}
{\mcitedefaultendpunct}{\mcitedefaultseppunct}\relax
\EndOfBibitem
\bibitem[{\citenamefont{Dobbs et~al.}(1992)\citenamefont{Dobbs, Darbellay, and
  Yeomans}}]{Dobbs1}
\bibinfo{author}{\bibfnamefont{H.~T.} \bibnamefont{Dobbs}},
  \bibinfo{author}{\bibfnamefont{G.~A.} \bibnamefont{Darbellay}},
  \bibnamefont{and} \bibinfo{author}{\bibfnamefont{J.~M.}
  \bibnamefont{Yeomans}}, \bibinfo{journal}{Europhys. Lett.}
  \textbf{\bibinfo{volume}{18}}, \bibinfo{pages}{439}
  (\bibinfo{year}{1992})\relax
\mciteBstWouldAddEndPuncttrue
\mciteSetBstMidEndSepPunct{\mcitedefaultmidpunct}
{\mcitedefaultendpunct}{\mcitedefaultseppunct}\relax
\EndOfBibitem
\bibitem[{\citenamefont{Dobbs and Yeomans}(1992)}]{Dobbs2}
\bibinfo{author}{\bibfnamefont{H.~T.} \bibnamefont{Dobbs}} \bibnamefont{and}
  \bibinfo{author}{\bibfnamefont{J.~M.} \bibnamefont{Yeomans}},
  \bibinfo{journal}{J. Phys.: Condens. Matter} \textbf{\bibinfo{volume}{4}},
  \bibinfo{pages}{10133} (\bibinfo{year}{1992})\relax
\mciteBstWouldAddEndPuncttrue
\mciteSetBstMidEndSepPunct{\mcitedefaultmidpunct}
{\mcitedefaultendpunct}{\mcitedefaultseppunct}\relax
\EndOfBibitem
\bibitem[{\citenamefont{Bauer and Dietrich}(2000)}]{Bauer1}
\bibinfo{author}{\bibfnamefont{C.}~\bibnamefont{Bauer}} \bibnamefont{and}
  \bibinfo{author}{\bibfnamefont{S.}~\bibnamefont{Dietrich}},
  \bibinfo{journal}{Phys. Rev. E} \textbf{\bibinfo{volume}{62}},
  \bibinfo{pages}{2428} (\bibinfo{year}{2000})\relax
\mciteBstWouldAddEndPuncttrue
\mciteSetBstMidEndSepPunct{\mcitedefaultmidpunct}
{\mcitedefaultendpunct}{\mcitedefaultseppunct}\relax
\EndOfBibitem
\bibitem[{\citenamefont{Osborn and Yeomans}(1995)}]{Osborn1}
\bibinfo{author}{\bibfnamefont{W.~R.} \bibnamefont{Osborn}} \bibnamefont{and}
  \bibinfo{author}{\bibfnamefont{J.~M.} \bibnamefont{Yeomans}},
  \bibinfo{journal}{Phys. Rev. E} \textbf{\bibinfo{volume}{51}},
  \bibinfo{pages}{2053} (\bibinfo{year}{1995})\relax
\mciteBstWouldAddEndPuncttrue
\mciteSetBstMidEndSepPunct{\mcitedefaultmidpunct}
{\mcitedefaultendpunct}{\mcitedefaultseppunct}\relax
\EndOfBibitem
\bibitem[{\citenamefont{Tolman}(1949)}]{Tolman1}
\bibinfo{author}{\bibfnamefont{R.~C.} \bibnamefont{Tolman}},
  \bibinfo{journal}{J. Chem. Phys.} \textbf{\bibinfo{volume}{17}},
  \bibinfo{pages}{333} (\bibinfo{year}{1949})\relax
\mciteBstWouldAddEndPuncttrue
\mciteSetBstMidEndSepPunct{\mcitedefaultmidpunct}
{\mcitedefaultendpunct}{\mcitedefaultseppunct}\relax
\EndOfBibitem
\bibitem[{\citenamefont{Blokhuis et~al.}(1999)\citenamefont{Blokhuis,
  Groenewold, and Bedeaux}}]{Blokhuis1}
\bibinfo{author}{\bibfnamefont{E.~M.} \bibnamefont{Blokhuis}},
  \bibinfo{author}{\bibfnamefont{J.}~\bibnamefont{Groenewold}},
  \bibnamefont{and} \bibinfo{author}{\bibfnamefont{D.}~\bibnamefont{Bedeaux}},
  \bibinfo{journal}{Mol. Phys.} \textbf{\bibinfo{volume}{96}},
  \bibinfo{pages}{397} (\bibinfo{year}{1999})\relax
\mciteBstWouldAddEndPuncttrue
\mciteSetBstMidEndSepPunct{\mcitedefaultmidpunct}
{\mcitedefaultendpunct}{\mcitedefaultseppunct}\relax
\EndOfBibitem
\bibitem[{\citenamefont{Helfrich}(1973)}]{Helfrich1}
\bibinfo{author}{\bibfnamefont{W.}~\bibnamefont{Helfrich}},
  \bibinfo{journal}{Z. Naturforsch. C} \textbf{\bibinfo{volume}{28}},
  \bibinfo{pages}{693} (\bibinfo{year}{1973})\relax
\mciteBstWouldAddEndPuncttrue
\mciteSetBstMidEndSepPunct{\mcitedefaultmidpunct}
{\mcitedefaultendpunct}{\mcitedefaultseppunct}\relax
\EndOfBibitem
\bibitem[{\citenamefont{Evans}(1979)}]{Evans2}
\bibinfo{author}{\bibfnamefont{R.}~\bibnamefont{Evans}}, \bibinfo{journal}{Adv.
  Phys.} \textbf{\bibinfo{volume}{28}}, \bibinfo{pages}{143}
  (\bibinfo{year}{1979})\relax
\mciteBstWouldAddEndPuncttrue
\mciteSetBstMidEndSepPunct{\mcitedefaultmidpunct}
{\mcitedefaultendpunct}{\mcitedefaultseppunct}\relax
\EndOfBibitem
\bibitem[{\citenamefont{Napi\'orkowski and Dietrich}(1993)}]{Napiorkowski1}
\bibinfo{author}{\bibfnamefont{M.}~\bibnamefont{Napi\'orkowski}}
  \bibnamefont{and} \bibinfo{author}{\bibfnamefont{S.}~\bibnamefont{Dietrich}},
  \bibinfo{journal}{Phys. Rev. E} \textbf{\bibinfo{volume}{47}},
  \bibinfo{pages}{1836} (\bibinfo{year}{1993})\relax
\mciteBstWouldAddEndPuncttrue
\mciteSetBstMidEndSepPunct{\mcitedefaultmidpunct}
{\mcitedefaultendpunct}{\mcitedefaultseppunct}\relax
\EndOfBibitem
\bibitem[{\citenamefont{Napi\'orkowski}(1994)}]{Napiorkowski2}
\bibinfo{author}{\bibfnamefont{M.}~\bibnamefont{Napi\'orkowski}},
  \bibinfo{journal}{Ber. Bunsenges. Phys. Chem.} \textbf{\bibinfo{volume}{98}},
  \bibinfo{pages}{352} (\bibinfo{year}{1994})\relax
\mciteBstWouldAddEndPuncttrue
\mciteSetBstMidEndSepPunct{\mcitedefaultmidpunct}
{\mcitedefaultendpunct}{\mcitedefaultseppunct}\relax
\EndOfBibitem
\bibitem[{\citenamefont{Napi\'orkowski and Dietrich}(1995)}]{Napiorkowski3}
\bibinfo{author}{\bibfnamefont{M.}~\bibnamefont{Napi\'orkowski}}
  \bibnamefont{and} \bibinfo{author}{\bibfnamefont{S.}~\bibnamefont{Dietrich}},
  \bibinfo{journal}{Z. Phys. B} \textbf{\bibinfo{volume}{97}},
  \bibinfo{pages}{511} (\bibinfo{year}{1995})\relax
\mciteBstWouldAddEndPuncttrue
\mciteSetBstMidEndSepPunct{\mcitedefaultmidpunct}
{\mcitedefaultendpunct}{\mcitedefaultseppunct}\relax
\EndOfBibitem
\bibitem[{\citenamefont{Bieker and Dietrich}(1998)}]{Bieker1}
\bibinfo{author}{\bibfnamefont{T.}~\bibnamefont{Bieker}} \bibnamefont{and}
  \bibinfo{author}{\bibfnamefont{S.}~\bibnamefont{Dietrich}},
  \bibinfo{journal}{Physica A} \textbf{\bibinfo{volume}{252}},
  \bibinfo{pages}{85} (\bibinfo{year}{1998})\relax
\mciteBstWouldAddEndPuncttrue
\mciteSetBstMidEndSepPunct{\mcitedefaultmidpunct}
{\mcitedefaultendpunct}{\mcitedefaultseppunct}\relax
\EndOfBibitem
\bibitem[{\citenamefont{Robledo et~al.}(1991)\citenamefont{Robledo, Varea, and
  Romero-Rochin}}]{Robledo1}
\bibinfo{author}{\bibfnamefont{A.}~\bibnamefont{Robledo}},
  \bibinfo{author}{\bibfnamefont{C.}~\bibnamefont{Varea}}, \bibnamefont{and}
  \bibinfo{author}{\bibfnamefont{V.}~\bibnamefont{Romero-Rochin}},
  \bibinfo{journal}{Physica A} \textbf{\bibinfo{volume}{177}},
  \bibinfo{pages}{474} (\bibinfo{year}{1991})\relax
\mciteBstWouldAddEndPuncttrue
\mciteSetBstMidEndSepPunct{\mcitedefaultmidpunct}
{\mcitedefaultendpunct}{\mcitedefaultseppunct}\relax
\EndOfBibitem
\bibitem[{\citenamefont{Romero-Rochin et~al.}(1991)\citenamefont{Romero-Rochin,
  Varea, and Robledo}}]{Rochin2}
\bibinfo{author}{\bibfnamefont{V.}~\bibnamefont{Romero-Rochin}},
  \bibinfo{author}{\bibfnamefont{C.}~\bibnamefont{Varea}}, \bibnamefont{and}
  \bibinfo{author}{\bibfnamefont{A.}~\bibnamefont{Robledo}},
  \bibinfo{journal}{Phys. Rev. A} \textbf{\bibinfo{volume}{44}},
  \bibinfo{pages}{8417} (\bibinfo{year}{1991})\relax
\mciteBstWouldAddEndPuncttrue
\mciteSetBstMidEndSepPunct{\mcitedefaultmidpunct}
{\mcitedefaultendpunct}{\mcitedefaultseppunct}\relax
\EndOfBibitem
\bibitem[{\citenamefont{Blokhuis and Bedeaux}(1992)}]{Blokhuis3}
\bibinfo{author}{\bibfnamefont{E.~M.} \bibnamefont{Blokhuis}} \bibnamefont{and}
  \bibinfo{author}{\bibfnamefont{D.}~\bibnamefont{Bedeaux}},
  \bibinfo{journal}{J. Comp. Phys.} \textbf{\bibinfo{volume}{97}},
  \bibinfo{pages}{3576} (\bibinfo{year}{1992})\relax
\mciteBstWouldAddEndPuncttrue
\mciteSetBstMidEndSepPunct{\mcitedefaultmidpunct}
{\mcitedefaultendpunct}{\mcitedefaultseppunct}\relax
\EndOfBibitem
\bibitem[{\citenamefont{Derjaguin}(1934)}]{Derjaguin1}
\bibinfo{author}{\bibfnamefont{B.}~\bibnamefont{Derjaguin}},
  \bibinfo{journal}{Kolloid Z.} \textbf{\bibinfo{volume}{69}},
  \bibinfo{pages}{155} (\bibinfo{year}{1934})\relax
\mciteBstWouldAddEndPuncttrue
\mciteSetBstMidEndSepPunct{\mcitedefaultmidpunct}
{\mcitedefaultendpunct}{\mcitedefaultseppunct}\relax
\EndOfBibitem
\bibitem[{\citenamefont{Goetzelmann and Dietrich}(1998)}]{Goetzelmann1}
\bibinfo{author}{\bibfnamefont{B.}~\bibnamefont{Goetzelmann}} \bibnamefont{and}
  \bibinfo{author}{\bibfnamefont{S.}~\bibnamefont{Dietrich}},
  \bibinfo{journal}{Phys. Rev. E} \textbf{\bibinfo{volume}{57}},
  \bibinfo{pages}{6785} (\bibinfo{year}{1998})\relax
\mciteBstWouldAddEndPuncttrue
\mciteSetBstMidEndSepPunct{\mcitedefaultmidpunct}
{\mcitedefaultendpunct}{\mcitedefaultseppunct}\relax
\EndOfBibitem
\bibitem[{\citenamefont{Troendle et~al.}(2008)\citenamefont{Troendle, Harnau,
  and Dietrich}}]{Troendle1}
\bibinfo{author}{\bibfnamefont{M.}~\bibnamefont{Troendle}},
  \bibinfo{author}{\bibfnamefont{L.}~\bibnamefont{Harnau}}, \bibnamefont{and}
  \bibinfo{author}{\bibfnamefont{S.}~\bibnamefont{Dietrich}},
  \bibinfo{journal}{J. Chem. Phys.} \textbf{\bibinfo{volume}{129}},
  \bibinfo{pages}{124716} (\bibinfo{year}{2008})\relax
\mciteBstWouldAddEndPuncttrue
\mciteSetBstMidEndSepPunct{\mcitedefaultmidpunct}
{\mcitedefaultendpunct}{\mcitedefaultseppunct}\relax
\EndOfBibitem
\bibitem[{\citenamefont{Abramowitz and Stegun}(1974)}]{Abramowitz1}
\bibinfo{author}{\bibfnamefont{M.}~\bibnamefont{Abramowitz}} \bibnamefont{and}
  \bibinfo{author}{\bibfnamefont{I.~A.} \bibnamefont{Stegun}},
  \emph{\bibinfo{title}{Handbook of Mathematical Functions}}
  (\bibinfo{publisher}{Courier Dover Publications}, \bibinfo{address}{New
  York}, \bibinfo{year}{1974})\relax
\mciteBstWouldAddEndPuncttrue
\mciteSetBstMidEndSepPunct{\mcitedefaultmidpunct}
{\mcitedefaultendpunct}{\mcitedefaultseppunct}\relax
\EndOfBibitem
\bibitem[{\citenamefont{Diehl et~al.}(1980)\citenamefont{Diehl, Kroll, and
  Wagner}}]{Diehl1}
\bibinfo{author}{\bibfnamefont{H.~W.} \bibnamefont{Diehl}},
  \bibinfo{author}{\bibfnamefont{D.~M.} \bibnamefont{Kroll}}, \bibnamefont{and}
  \bibinfo{author}{\bibfnamefont{H.}~\bibnamefont{Wagner}},
  \bibinfo{journal}{Z. Phys. B} \textbf{\bibinfo{volume}{36}},
  \bibinfo{pages}{329} (\bibinfo{year}{1980})\relax
\mciteBstWouldAddEndPuncttrue
\mciteSetBstMidEndSepPunct{\mcitedefaultmidpunct}
{\mcitedefaultendpunct}{\mcitedefaultseppunct}\relax
\EndOfBibitem
\bibitem[{\citenamefont{Kreyszig}(1991)}]{Kreyszig1}
\bibinfo{author}{\bibfnamefont{E.}~\bibnamefont{Kreyszig}}, in
  \emph{\bibinfo{booktitle}{Differential geometry}}
  (\bibinfo{publisher}{Courier Dover Publications}, \bibinfo{address}{New
  York}, \bibinfo{year}{1991})\relax
\mciteBstWouldAddEndPuncttrue
\mciteSetBstMidEndSepPunct{\mcitedefaultmidpunct}
{\mcitedefaultendpunct}{\mcitedefaultseppunct}\relax
\EndOfBibitem
\bibitem[{\citenamefont{Giessen et~al.}(1998)\citenamefont{Giessen, Blokhuis,
  and Bukman}}]{Giessen2}
\bibinfo{author}{\bibfnamefont{A.~E.} \bibnamefont{Giessen}},
  \bibinfo{author}{\bibfnamefont{E.~M.} \bibnamefont{Blokhuis}},
  \bibnamefont{and} \bibinfo{author}{\bibfnamefont{D.~J.}
  \bibnamefont{Bukman}}, \bibinfo{journal}{J. Chem. Phys.}
  \textbf{\bibinfo{volume}{108}}, \bibinfo{pages}{1148}
  (\bibinfo{year}{1998})\relax
\mciteBstWouldAddEndPuncttrue
\mciteSetBstMidEndSepPunct{\mcitedefaultmidpunct}
{\mcitedefaultendpunct}{\mcitedefaultseppunct}\relax
\EndOfBibitem
\bibitem[{\citenamefont{Segovia-Lopez and Romero-Rochin}(2006)}]{Segovia1}
\bibinfo{author}{\bibfnamefont{J.~G.} \bibnamefont{Segovia-Lopez}}
  \bibnamefont{and}
  \bibinfo{author}{\bibfnamefont{V.}~\bibnamefont{Romero-Rochin}},
  \bibinfo{journal}{Phys. Rev. E} \textbf{\bibinfo{volume}{73}},
  \bibinfo{pages}{021601} (\bibinfo{year}{2006})\relax
\mciteBstWouldAddEndPuncttrue
\mciteSetBstMidEndSepPunct{\mcitedefaultmidpunct}
{\mcitedefaultendpunct}{\mcitedefaultseppunct}\relax
\EndOfBibitem
\bibitem[{\citenamefont{Dean and Horgan}(2006)}]{Dean1}
\bibinfo{author}{\bibfnamefont{D.~S.} \bibnamefont{Dean}} \bibnamefont{and}
  \bibinfo{author}{\bibfnamefont{R.~R.} \bibnamefont{Horgan}},
  \bibinfo{journal}{Phys. Rev. E} \textbf{\bibinfo{volume}{73}},
  \bibinfo{pages}{011906} (\bibinfo{year}{2006})\relax
\mciteBstWouldAddEndPuncttrue
\mciteSetBstMidEndSepPunct{\mcitedefaultmidpunct}
{\mcitedefaultendpunct}{\mcitedefaultseppunct}\relax
\EndOfBibitem
\bibitem[{\citenamefont{Blokhuis and Bedeaux}(1991)}]{Blokhuis2}
\bibinfo{author}{\bibfnamefont{E.~M.} \bibnamefont{Blokhuis}} \bibnamefont{and}
  \bibinfo{author}{\bibfnamefont{D.}~\bibnamefont{Bedeaux}},
  \bibinfo{journal}{J. Comp. Phys.} \textbf{\bibinfo{volume}{95}},
  \bibinfo{pages}{6986} (\bibinfo{year}{1991})\relax
\mciteBstWouldAddEndPuncttrue
\mciteSetBstMidEndSepPunct{\mcitedefaultmidpunct}
{\mcitedefaultendpunct}{\mcitedefaultseppunct}\relax
\EndOfBibitem
\end{mcitethebibliography}

\end{document}